# Quantitative Determination of the Probability of Multiple-Motor Transport in Bead-Based Assays

Qiaochu Li[1], Stephen J. King[2], Ajay Gopinathan[1], and Jing Xu[1,*]

[1]Department of Physics, University of California, Merced, CA 95343, USA

[2]Burnett School of Biomedical Sciences, University of Central Florida, FL 32827, USA

[*]**Correspondence:** Jing Xu (jxu8@ucmerced.edu)

**ABSTRACT:**

With their longest dimension typically less than 100 nm, molecular motors are significantly below the optical-resolution limit. Despite substantial advances in fluorescence-based imaging methodologies, labeling with beads remains critical for optical trapping-based investigations of molecular motors. A key experimental challenge in bead-based assays is that the number of motors on a bead is not well defined. Particularly for single-molecule investigations, the probability of single- versus multiple-motor events has not been experimentally investigated. Here, we used bead travel distance as an indicator of multiple-motor transport and determined the lower-bound probability of bead transport by two or more motors. We limited the ATP concentration in order to increase our detection sensitivity for multiple- versus single-kinesin transport. Surprisingly, for all but the lowest motor number examined, our measurements exceeded estimations of a previous model by two or more fold. To bridge this apparent gap between theory and experiment, we derived a closed-form expression for the probability of bead transport by multiple motors, and constrained the only free parameter in this model using our experimental measurements. Our data indicate that kinesin extends to ~57 nm during bead transport, suggesting that kinesin exploits its conformational flexibility to interact with microtubules at highly curved interfaces such as those present for vesicle transport in cells. To our knowledge, our findings provide the first experimentally constrained guide for the probability of multiple-motor transport in optical trapping studies. The experimental approach utilized here (limiting ATP concentration) may be generally applicable to studies in which molecular motors are labeled with cargos that are artificial or are purified from cellular extracts.

# INTRODUCTION

Microtubule-based molecular motors are protein machines that drive long-range mechanical transport in cells (1). This transport process is critical for cellular function and survival. Quantitative understanding of molecular-motor function at the single-molecule level is crucial for deciphering the complexity of mechanical transport in cells and for engineering biomimetic machineries on the nano/micro scale. Optical trapping has been instrumental in elucidating the mechanochemical functions of molecular motors, particularly at the single-molecule level (2-4).

Perhaps paradoxically, the single-molecule range in optical trapping experiments has yet to be defined in quantitative terms. Uncertainty arises because optical trapping studies use dielectric beads to label molecular motors. Each bead is typically decorated by a Poisson-distributed number of motors, rather than by a well-defined number of motors. The three-dimensional nature of the bead further complicates the problem, since not all motors on the bead can reach the microtubule at the same time. While a reduction in the ratio of motors to beads can lower the probability of cargo transport by two or more motors, this reduction also limits the fraction of motor/bead complexes capable of interacting with the microtubule and in turn significantly reduces experimental throughput. Although force measurements can shed light on the number of motors engaged in transport, work by Furuta et al. (5) revealed that, depending on motor type, the force generated by multiple motors is not necessarily sensitive to motor number. Previously, Beeg et al. (6) determined the average number of simultaneously engaged motors on a bead for a range of motor concentrations, making it possible to evaluate the probability of multiple-motor events measured under the same experimental conditions. However, Beeg et al. (6) used a bead size (100 nm diameter) substantially smaller than that optimized for optical trapping (~500 nm diameter, (4)). Because a 5-fold increase in bead size can impact both the density of motors on the bead surface and the fraction of bead surface available to the microtubule, it is non-trivial to translate the experimental conditions from (6) to those suitable for optical trapping.

To date, only one theory model has provided a general guide for estimating the probability of multiple-motor transport at a given motor number in bead-based assays (3). However, the assumptions of this previous model have not been experimentally verified. Importantly, the motors are assumed to be fully extended and their motor domains effectively in contact with each other (Fig. S1 *A* in the Supporting Material). There is increasing consensus that motors should be able to bind different locations along the length of the microtubule (7, 8) (Fig. S1 *B*). However, this updated geometry does not readily lend itself to theory predictions, since there is limited information on a key parameter in this updated geometry (how far the motor can extend during bead transport, Fig. S1 *B*). As a result, predictions from the previous model (3) have remained the only guide for estimating the probability of multiple-motor transport in optical trapping studies (9-13).

Here we sought to develop a quantitative guide to understanding the probability of multiple-motor events in the dilute motor range. We focused on experimental details that are typically used in optical trapping studies (500 nm diameter spherical beads and randomly distributed motors on the bead surface) and carried out our experiments using the major microtubule-based motor, kinesin-1 (conventional kinesin). We used motile fraction, the probability of beads exhibiting motility along microtubules, as an experimental readout for the average number of active motors on the bead (2, 3) (Supporting Text 1 and Fig. S2). We used bead travel distance to identify bead transport by multiple kinesins for a range of motile fractions, and used a limiting

ATP concentration to increase our detection sensitivity for these multiple-motor events (14-18). Surprisingly, for all but the lowest motile fractions examined, we detected substantially higher probabilities of multiple-motor transport than that predicted by the previous theory model (3). We therefore applied both theory and simulation to unravel this discrepancy.

## MATERIALS AND METHODS

### Proteins and reagents

Bovine brain tubulin was purified over a phosphocellulose column as previously described (19). Conventional kinesin was purified from bovine brain as previously described (20), except that 9S kinesin was eluted from the Mono-Q resin using customized salt gradients to separate kinesin from other polypeptides in the 9S sucrose fractions (21). Chemicals were purchased from Sigma Aldrich (St. Louis, MO, USA).

### Microtubule preparation

To assemble taxol-stabilized microtubules, purified tubulin (40 µM) was supplemented with 0.5 mM GTP and incubated for 20 min at 37 °C. Assembled microtubules were mixed with an equal volume of PM buffer (100 mM PIPES, 1 mM $MgSO_4$, 2 mM EGTA, pH 6.9) supplemented with 40 µM taxol and incubated for 20 min at 37 °C. Microtubules were then kept at room temperature in a dark box and used within four days of preparation.

### Motor/bead complex preparation

Kinesin was incubated with carboxylated polystyrene beads (500 nm diameter, Polysciences, Warrington, PA, USA) in motility buffer (67 mM PIPES, 50 mM $CH_3CO_2K$, 3 mM $MgSO_4$, 1 mM dithiothreitol, 0.84 mM EGTA, 10 µM taxol, pH 6.9) for 10 min at room temperature; this solution was supplemented with an oxygen-scavenging solution (250 µg/mL glucose oxidase, 30 µg/mL catalase, 4.6 mg/mL glucose) and ATP (1 mM or 0.01 mM, as indicated) prior to motility measurements. Bead concentration was kept constant at $3.6\times10^5$ particles/µL and the concentration of kinesin was varied to give rise to a range of motile fractions (Fig. S2).

### In vitro optical trapping

Optical trapping was carried out in flow cells, imaged via differential interference microscopy, and video-recorded at 30 Hz as previously described (16). For all studies here, we limited the trap power to <20 mW (at fiber output), such that the trap positioned individual beads but was not sufficient to stall beads carried by a single kinesin (stall force ~4.5 pN, (22)). To measure motile fraction, we used the optical trap to position individual beads in the vicinity of the microtubule; a motile event was scored if and only if the bead demonstrated directed motion away from the center of the trap. We observed occasional events where the bead appeared to bind (as indicated by reduced Brownian motion of the bead) but did not move processively away from the trap center (data not shown). We did not score these events for our motile fraction measurements, as they constituted neither motile events nor non-motile events. The beads in these events typically detached shortly after we turned off the optical trap, without demonstrating any clear movement along the microtubule (data not shown). For motile beads, upon observation of directed bead motion along the microtubule, we turned off the optical trap in order to enable cargo transport in the absence of external load.

**Data analysis**

Video recordings of bead motion were particle-tracked to 10 nm resolution (1/3 pixel) using a template-matching algorithm as previously described (16, 23). Travel distance for each bead was determined as the net displacement of the bead along the microtubule axis upon the bead's binding to and then detaching from the microtubule. The distribution of travel distances for each experimental condition (motile fraction and ATP concentration) was fitted to a single-exponential decay (2). Mean travel distance and the associated standard error of the mean for each distribution were determined from the fitted decay constant and uncertainty, respectively. To account for human reaction time to manually shut off the optical trap, only trajectories >0.3 µm were used to determine the distribution of travel distances at each motile fraction.

Best-fit of motile fraction measurements to a one-motor Poisson curve was carried out in OriginPro9.1 (OriginLab Corp., Northampton, MA, USA). Least $\chi^2$ fitting was used to constrain our theory model using our experimental measurements, via a custom routine in MatLab (The MathWorks, Inc., Natick, MA, USA).

**Simulation**

Our simulation randomly distributes a mean number of active motors ($n$) on a 51x51 lattice with periodic boundaries (a torus lattice), identifies the location of a lattice site that is occupied by one motor, and evaluates the number of motors in a closed patch surrounding this particular lattice site. The size of the closed path is determined as the area of the square lattice multiplied by $\alpha$, the probability that any two randomly distributed motors on the bead are in close enough proximity to be in simultaneous reach of the microtubule. We repeated the simulation 10,000 times to determine the probability of counting two or more motors for each set of $n$ and $\alpha$ tested. We evaluated the motile fraction corresponding to each $n$ using the relationship *motile fraction* $= 1 - e^{-n}$ (Fig. S2 and (2, 3)).

**Statistical analysis**

Standard errors for binomially distributed data were calculated as $\sqrt{P(1-P)/N}$, where $P$ is the measured fraction and $N$ is the sample size. We did not observe a substantial difference between this error calculation and that using a 68.3% confidence interval for a binomial distribution (data not shown). In case of a zero-fraction measurement (at motile fraction = 0.2, Fig. 1 *B*), the standard error was estimated as the upper level of the 68.3% confidence interval for a binomial distribution.

## RESULTS

**Lower-bound measurements of the probability of multiple-motor transport**

We used bead travel distance to identify multiple-motor events versus single-motor events. The average travel distance for a single kinesin is ~1 μm (2, 24). When a bead is transported by a single kinesin, there is a <0.1% probability that the travel distance will be >6.9 μm (Supporting Text 2). Thus, when we observed a motile event longer than our travel threshold, we excluded it from the set of single-motor events with high confidence and identified it as a multiple-motor event. Since not all multiple-motor events exceed this 6.9 μm threshold, our measurements of long-travel events represent lower-bound values for the probability of multiple-motor transport.

We used ATP as an experimental handle to increase the likelihood that a multiple-motor event travels >6.9 μm (blue bars, Fig. 1 *A*). Previous studies demonstrated that the distance traveled by multiple kinesins is inversely tuned by ATP concentration (0.01-1 mM) (14-18). Importantly, the single-kinesin travel distance remains constant over the same ATP range (15, 24). We thus anticipated that a limiting ATP concentration (0.01 mM) would increase our detection sensitivity for multiple-motor events.

We varied the input kinesin concentration to experimentally access a range of motile fractions (or motor numbers) (Fig. S2 *A*). For each motile fraction, we measured 50-150 motile trajectories and determined the corresponding distribution of travel distances (Fig. 1 *A*). For both ATP concentrations examined, we found that the fraction of long-travel events increased with increasing motile fraction (blue bars, Fig. 1 *A*). At the lowest motile fraction tested (0.2), we did not observe any long-travel events for either ATP concentration. At the highest motile fraction tested (0.8), the fraction of long-travel events was substantially larger than zero for both 0.01 mM and 1 mM ATP. This observation is consistent with the increased presence of motors on the bead for the higher motile fractions (Fig. S2 *B*).

Importantly, for the higher motile fractions (>0.2), we observed substantially higher fractions of long-travel events at the lower ATP concentration ($F_{>6.9\mu m}$, Fig. 1 *A*). For example, at a motile fraction of 0.8, the fraction of long-travel events was 23% at 0.01 mM ATP, approximately 4-fold larger than the 6% at 1 mM ATP. Since we used the same trap power for measurements at both ATP concentrations, the increase in the fraction of long-travel events was due to the change in ATP concentration, not experiment artifacts from optical trapping. Further, for measurements below our travel threshold (grey shading, Fig. 1 *A*), the mean travel distances measured at 1 mM ATP approximately doubled as the motile fraction increased from 0.2 to 0.8, while the mean travel distances measured at 0.01 mM ATP remained approximately constant at the single-kinesin level (~1 μm, (2, 24)) over the same motile fraction range. This observation indicates that only the travel distance, but not the likelihood, of multiple-motor events was amplified by the lower ATP concentration. Given the increase in multiple-motor travel distance at 0.01 mM ATP, multiple-motor events are less likely to influence the distribution of travel distances shorter than 6.9 μm. Together, these data demonstrate that our approach of limiting ATP is effective in uncovering the presence of multiple-motor transport in bead-based assays.

We measured the fraction of long-travel events for a range of motile fractions at 0.01 mM ATP (blue scatters, Fig. 1 *B*). We detected a substantial difference between these lower-bound measurements and predictions of a previous model (3) (Supporting Text 3) (blue scatters vs. magenta line, Fig. 1 *B*). For motile fractions >0.4, our lower-bound measurements were substantially larger than predictions from the previous model. For example, at a motile fraction of 0.5, our lower-bound measurements indicate that at least 8% of motile trajectories are transported by multiple motors, more than twice the 3% predicted previously.

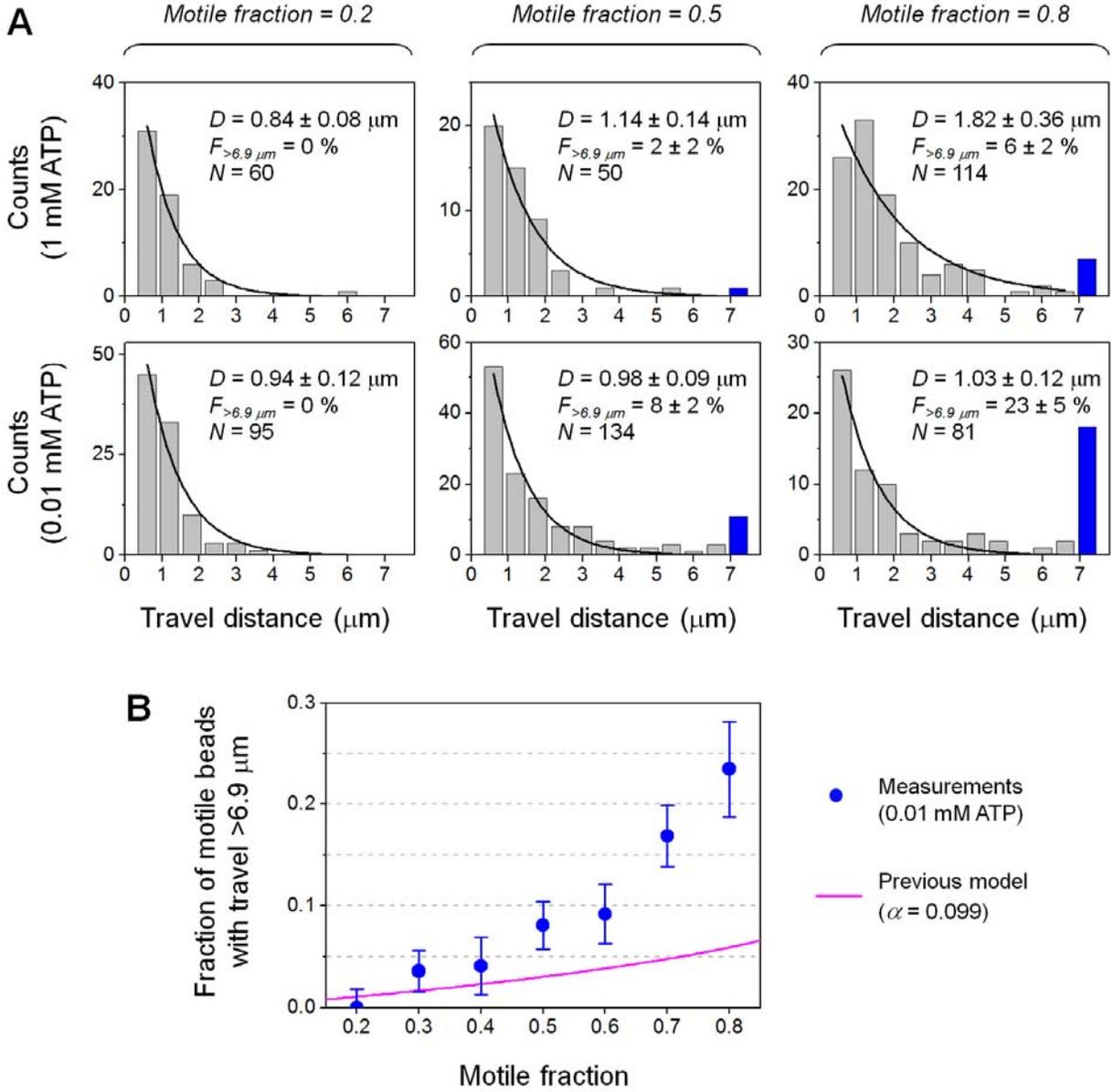

FIGURE 1   Measurements of bead travel distances for different motile fractions. (*A*) Distribution of travel distances for three motile fractions measured at 1 mM ATP (top) and 0.01 mM ATP (bottom). Solid line, best fit to a single exponential decay for travel distances ≤6.9 μm. Blue bar, cumulative counts of events with travel distance >6.9 μm. $D$ (± standard error), fitted mean travel distance for measurements shorter than our travel threshold. $F_{>6.9\mu m}$ (± standard error), fraction of motile events longer than our travel threshold. $N$, sample size. (*B*) Fraction of motile beads with travel >6.9 μm,

measured at 0.01 mM ATP (blue scatter). Error bars, standard error. Magenta line, predicted fraction of motile events transported by two or more motors, based on a previous theory (Supporting Text 3, (3)).

We examined the possibility that our measurements were influenced by application of the optical trap's force to the motors. In order to measure bead travel distance, we used an optical trap to initially confine individual beads to the vicinity of a microtubule. Previous studies reported that the force exerted by an optical trap can promote synergistic cooperation between multiple kinesins (25, 26). It is therefore possible that our use of an optical trap may have biased multiple-motor travel distance toward higher values by influencing how many motors are engaged in transport at the start of a multiple-motor event (before the trap is turned off). We speculate that the magnitude of this effect in our experiments is small, since the trap's effect is more pronounced when the trap's force is larger than or comparable to the force produced by a single kinesin (25, 26). In the current study, we used a trap with substantially lower power; beads carried by ~one motor moved processively away from the trap center without detaching prematurely (for example, motile fraction = 0.2, Fig. 1 *A*), which is difficult to achieve when the trap's force is comparable to the motor's force (26-28). Importantly, regardless of the magnitude of the trap's effect on multiple-motor travel distance, the trap's force cannot increase the distance traveled by a single motor under any circumstances (26-28). Thus, while the trap's force may improve the sensitivity for detecting a multiple-motor event, it cannot lead to false identification of a single-motor event as a multiple-motor event.

To understand the discrepancy between our measurements and the previous model (Fig. 1 *B*), we next derived a general theory model for the probability of transport by multiple motors (Eqs. 1 and 2, and Fig. 2). We then used our lower-bound measurements to constrain the free parameter in our model and to gain insight into the geometry of multiple-motor transport in bead-based assays (Fig. 3).

**An exact theory expression for the probability of multiple-motor transport**

Using the same framework as developed in the previous model (3), we described the probability of bead transport by two or more motors as the weighted sum $P(\geq 2) = \sum_{k=2}^{\infty} p(k|n) \cdot g(2|k)$, where $n$ is the average number of motors on the bead, $p(k|n)$ is the Poisson probability that there are exactly $k$ motors on the bead, and $g(2|k)$ is the probability that at least two of the $k$ motors on the bead are available for transport. Whereas the Poisson probability $p(k|n) = n^k e^{-n}/k!$ only concerns the total number of motors on the bead and does not differ between experiments, the weighting factor $g(2|k)$ is sensitive to experimental details (bead size) and assumptions of multiple-motor transport geometry (Fig. S1). As detailed below, the previous model (magenta line, Fig. 1 *B*) used an approximation for this weighting factor. We hypothesized that incorporating the exact expression of $g(2|k)$ would help bridge the gap between experiment and theory in Figure 1 *B*.

To arrive at an exact expression for $g(2|k)$, we used $\alpha$ to denote the probability that any two randomly attached motors are within simultaneous reach of the microtubule. The value of $\alpha$ is important, since it critically impacts the probability of multiple-motor transport for a given motor number *n*. Intuitively, the larger the value of $\alpha$, the more likely it is for two or more motors to engage in transport at a given motor number. The evaluation of $\alpha$ is difficult, since it depends sensitively bead size, as well as assumptions of two-motor transport geometry (Fig. S1). Experimental measurements in the current study (Fig. 1 *B*) provided the first opportunity to constrain the value of $\alpha$ for kinesin without any geometry assumptions.

The exact expression for the weighting factor is $g(2|k) = 1-(1-\alpha)^{k-1}$. For any one motor present on the bead, the probability that none of the remaining $k-1$ motors are within simultaneous reach of the microtubule is $(1-\alpha)^{k-1}$. This exact expression differs from the approximated form, $g(2|k) = \alpha$, used in a previous model (3). The previous approximation ($g(2|k) = \alpha$) underestimates the weighting factor for all $k > 2$. For example, when there are 20 motors on the bead, the probability that two or more motors are within simultaneous reach of the microtubule is larger than that when only two motors are present on the bead.

Using $g(2|k) = 1-(1-\alpha)^{k-1}$, we arrived at a closed-form description for the probability of multiple-motor transport (detailed derivation in Supporting Text 4):

$$P(\geq 2) = 1 + e^{-n}\left(\frac{\alpha}{1-\alpha}\right) - \frac{e^{-n\alpha}}{1-\alpha},\qquad\text{Eq. 1}$$

where *n* indicates the average number of active motors present on the bead. Since we absorbed geometry considerations into the probability $\alpha$, this closed-form expression (Eq. 1) is general and does not depend on details of multiple-motor geometry (for example, Fig. S1 *A* vs. S1 *B*). We then recast the dependence on the mean motor number (*n*) as that on the experimental measurable, motile fraction, using the relationship *motile fraction* $= 1-e^{-n}$ (Fig. S2 and (2, 3)).

We used a numerical simulation to test the validity of Equation 1 (Fig. 2). Our numerical simulation captured the three-dimensional nature of the cargo via the usage of a torus lattice; we also included one free parameter to reflect the probability $\alpha$ (Materials and Methods). Predictions from Equation 1 were in excellent agreement with the results of our numerical simulations for all $\alpha$ values tested (black line vs. scatter, Fig. 2). On the other hand, when we approximated the probability of multiple-motor transport using $g(2|k) = \alpha$ as in the previous model, we consistently obtained a lower value than that returned by our numerical simulation (magenta line vs. black scatter, Fig. 2). The extent of underestimation was more pronounced for small values of $\alpha$ and diminished as $\alpha$ approached 1. This scenario is expected, since in the limit of $\alpha = 1$, both the exact and the approximate forms of the weighting factor also converge to 1. These data support our hypothesis that the underestimation in $g(2|k)$ contributes to the gap between experiment and theory in Figure 1 *B*.

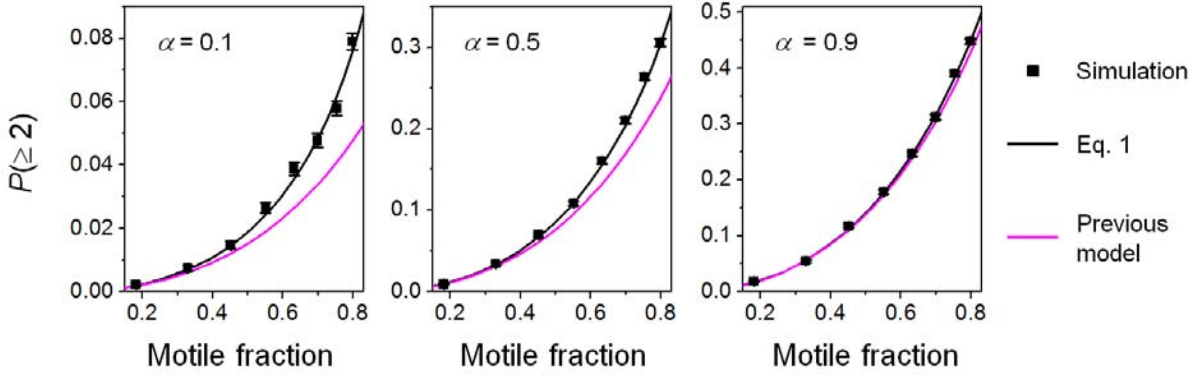

FIGURE 2   Probability that a bead is transported by two or more motors, $P(\geq 2)$, evaluated for three values of $\alpha$ (the probability that two randomly attached motors are within simultaneous reach of the microtubule). Error bars in simulation results indicate standard error (sample size $N$ = 10,000 per simulation condition).

Interestingly, correcting $g(2|k)$ alone was not sufficient to bridge the gap between experiment and theory. When evaluated using the previously estimated value for $\alpha$ (0.099), the probability of multiple-motor transport (as determined by Eq. 1 and simulation, black line and scatters, Fig. 2) remained substantially smaller than our lower-bound measurements (blue scatters, Fig. 1 *B*). For example, at a motile fraction of 0.8, we anticipate that ~10% of the motile events are transported by two or more motors ($\alpha$ = 0.1, Fig. 2). This value is less than half of the 23% observed experimentally (Fig. 1). Thus, our theory study indicates that the value of $\alpha$ is substantially higher than previously anticipated. This finding is perhaps not surprising, since the geometry of two-motor transport used in the previous estimation was itself an approximation (Fig. S1 *A*). Importantly, in the updated geometry for two-motor transport, the value of $\alpha$ becomes larger than its previous estimated value (0.099) when kinesin extends beyond 16% of its contour length (80 nm, (29, 30)) (Fig. S1 *B*). Since kinesin has been found to extend to at least 28% of its contour length during active transport (31), it is conceivable that $\alpha$ may be larger than previously estimated.

**Quantitative comparison between experiment and theory**

To enable direct comparison between theory and experiments, we recast Equation 1 to reflect the lower-bound nature of our experimental measurements. Our experimental measurements represent lower-bound values, since our distance threshold excludes the population of multiple-motor transport events that travel ≤6.9 μm (Fig. 1). Imposing the same threshold, we derived (Supporting Text 5) the expected lower-bound probability of multiple-motor transport as

$$P_{lower-bound}(\geq 2) = P(\geq 2) - f \cdot \left( \frac{\alpha \cdot e^{-n\alpha}}{(1-\alpha)^2} - \frac{\alpha \cdot n \cdot e^{-n}}{1-\alpha} - \frac{\alpha \cdot e^{-n}}{(1-\alpha)^2} \right), \qquad \text{Eq. 2}$$

where $P(\geq 2)$ is as described in Equation 1 and $f = 0.556\pm0.096$, as we previously measured for the experimental condition used in the current investigation (0.01 mM ATP) (15). Similar to Equation 1, this description is a function of the experimental measureable motile fraction and the free parameter, $\alpha$.

Constraining Equation 2 using our experimental measurements, we obtained a best-fit value of $\alpha = 0.405$ (Fig. 3 A). Note that since our theory expressions describe the probability of multiple-motor transport for all beads (including those without active motors), we scaled our measurements in Figure 1 B by their associated motile fractions to obtain the lower-bound measurements on the same probability (blue scatters, Fig. 3 A). As this best-fit value of $\alpha$ is 4-fold larger than that estimated using the previous model (0.099, Supporting Text 3), our data indicate that the geometry assumed in the previous model (Fig. S1 A) is unlikely to occur during bead transport.

Using an updated two-motor geometry (7, 8) (Fig. S1 B) and our best-fit value of $\alpha = 0.405$ (Fig. 3 A), we arrived at a mean extension length of 57 nm for kinesin (Fig. 3 B). This mean extension length and its associated upper and lower limits (79 nm and 40 nm, respectively) are reasonable as they are within kinesin's contour length (80 nm (29, 30)). These data support the updated geometry and address the question of kinesin's extension length for the typical bead size used in optical trapping (500 nm diameter).

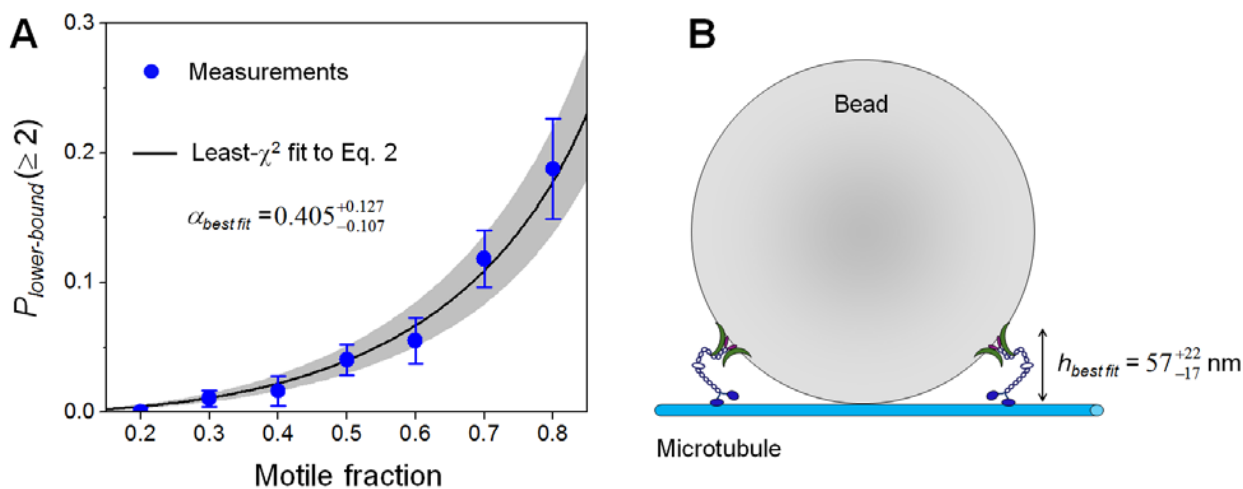

FIGURE 3  Quantitative comparison between experiment and theory. (A) The lower-bound probability of a bead transported by two or more motors. Scatter, experimental measurements. Error bars, standard error. Solid line and shaded area, least-$\chi^2$ fit and standard error, respectively, for Equation 2. (B) An updated geometry for two-motor transport (not to scale). $h_{best\ fit}$, extension length of the kinesin motor, corresponding to the best-fit value of $\alpha$ determined in (A).

### Estimated fraction of motile beads transported by multiple motors

Based on our findings (Fig. 3 and Eq. 1), we evaluated the probability that a motile bead is transported by two or more motors (Fig. 4). To do so, we normalized our expression in Equation 1 by the associated motile fraction. Since only motile events can contribute to transport measurements, the normalized probability directly reflects the impact of multiple-motor events on the measured motility parameters.

Using the best-fit $\alpha$ value for kinesin (Fig. 3), we evaluated the fraction of motile beads transported by multiple kinesins (Fig. 4, left). For all motile fractions measured, our experimentally constrained estimations were >4-fold larger than previous estimations (black line vs. magenta line, Fig. 4, left). We estimate that for motile fractions below 0.387, ≥90% of motile events are due to the action of a single kinesin (Fig. 4, left).

Using the updated two-motor geometry supported by our study (Fig. 3 *B*), we extended our evaluation to another major microtubule-based motor, dynein (Fig. 4, right). Here, we could not constrain the extension length of dynein as we did for kinesin, since there is currently no experimental handle to amplify the difference between single- and multiple-dynein travel distances. Instead, we referred to previous structural studies (32-34) to estimate the range for dynein's motor extension length. This approach is reasonable, since the flexibility of dynein is expected to be substantially more limited than that of kinesin (1).

We used the size of dynein's motor domain to place a lower bound on its extension length (26 nm (34)), and obtained a minimum $\alpha$ value of 0.197 (Fig. S1 *B*). This lower limit may be more relevant for studies using minimal dynein constructs containing only the motor domains. Estimations using this lower limit remained >5-fold larger than previous predictions for dynein (black dashed line vs. magenta line, Fig. 4, right). We estimate that the motile fraction can be as high as 0.609 while still ensuring that <10% of the motile events are transported by multiple dyneins (dashed black line, Fig. 4, right).

We used dynein's contour length as the maximum extension length (50 nm, (32, 33)), corresponding to a maximum $\alpha$ value of 0.36 (Fig. S1 *B*). As expected, the probability of multiple-motor transport increased at the higher $\alpha$ value. Using this upper-bound value, we estimate that <10% of motile measurements reflect transport by multiple dyneins for motile fractions below 0.420 (solid black line, Fig. 4, right).

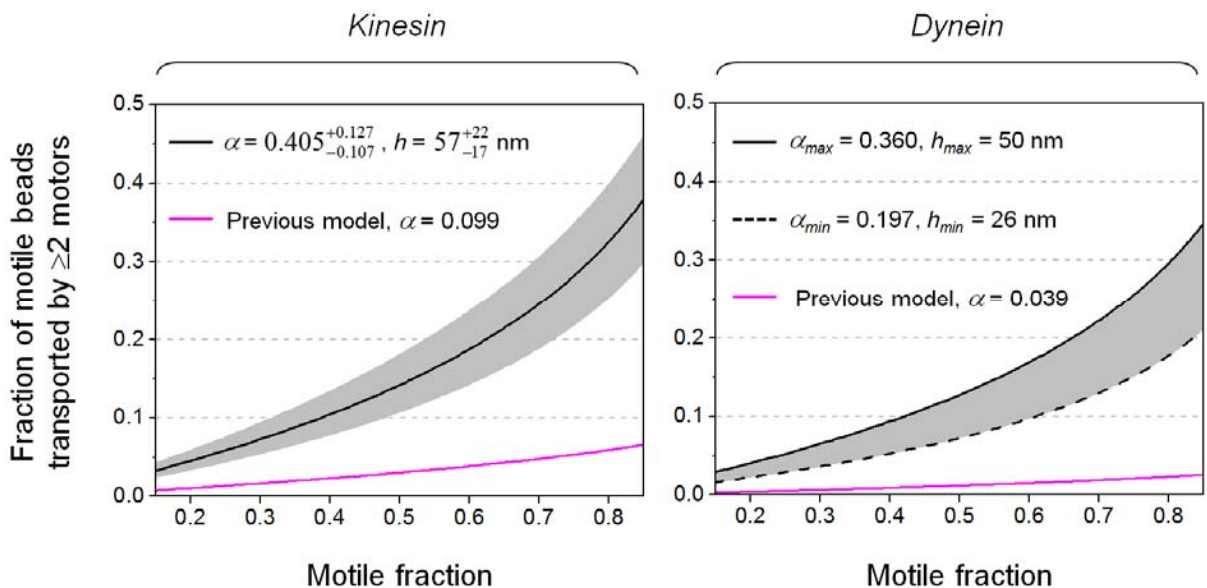

FIGURE 4  Fraction of motile beads transported by multiple motors as a function of motile fraction, estimated for two major microtubule-based molecular motors. Magenta lines, a previous model (3). Black lines and grey area, this study. $\alpha$, the probability that two randomly attached motors are within simultaneous reach of the microtubule. $h$, the extension length of the motor while bound to a bead. (Left) Solid black line and grey area, mean ± standard error based on our experimentally constrained $\alpha$ value for kinesin (Fig. 3). (Right) Dashed black line and solid black line, minimum and maximum values based on the updated two-motor geometry (Fig. S1 *B*) and the motor's extension length, $h$, supported by previous structural studies of dynein (32-34).

## DISCUSSION

Here, we combined experiments and theory to determine the probability of multiple-motor transport in bead-based assays. In the experimental portion of our study, we increased our detection sensitivity for multiple- versus single-kinesin transport using ATP concentration as an experimental handle to tune the multiple-kinesin travel distance (Fig. 1). We verified that the increased likelihood of long-travel events was directly associated with lower ATP levels (Fig. 1 *A*). In our theory work, we derived an exact and closed-form expression for the probability of multiple-motor transport that contains only one free parameter, $\alpha$ (Eq. 1). Our numerical simulation validated the predictions of our expression for a range of $\alpha$ values (Fig. 2). We then recast our expression to reflect the lower-bound nature of our experimental measurements (Eq. 2). We constrained the one free parameter, $\alpha$, in our theory model using our experimental measurements (Fig. 3 *A*). The resulting predictions constitute a set of quantitative guides for the probability of multiple-motor contributions in single-molecule investigations using optical trapping (Fig. 4). To our knowledge, this is the first set of such guides that has been experimentally constrained.

We focused the current study on experimental details used in typical optical trapping studies; specifically, individual motors were uniformly distributed atop spherical beads (500 nm diameter). Under these conditions, we found that it was not necessary to limit single-molecule investigations to motile fractions substantially below 0.2 (such as 0.07 in (9)), where <5% of motile events were due to multiple motors for both kinesin and dynein (Fig. 4). Our study also supports the use of a motile fraction of 0.3 for single-molecule investigations (10-13), since we estimated that 7±2% motile events were due to multiple kinesins (<7% for multiple dyneins; Fig. 4). Depending on the expected cooperativity of motors functioning in groups and specific requirements for measurement precision, experiments may be carried out at somewhat higher motile fractions to boost experimental throughput.

In general, our results in Figure 4 apply only to studies in which the motors are randomly distributed on the bead surface, since we assumed a single parameter, $\alpha$, across the bead surface. Despite this restriction, our model is appropriate for studies that use intermediate binding sites (antibodies (5, 15) and/or DNA scaffolds (5, 11)) to group motors together locally on the bead surface. In these special cases, the intermediate binding sites are randomly distributed atop the bead surface. Motors bound to individual intermediate sites may be considered to constitute a single "super-motor" complex; evaluation of Equation 1 yields the probability that a bead is transported by multiple super-motor complexes (11, 15).

Further studies are necessary to understand how changing the bead size impacts the predictions in Figure 4. Our study supports an updated geometry for two-motor transport in which the motors bind at different locations along the length of the microtubule (Fig. 3). In this updated geometry, the value of $\alpha$ depends sensitively on both the bead size and the motor extension length (Fig. S1 *B*). For a given extension length, the larger the bead, the less likely it will be for two randomly attached motors to be within simultaneous reach of the same microtubule, thus reducing $\alpha$. Therefore, if the motor is relatively rigid (as is likely the case for dynein (1)), an increase in bead size (for example, to the ~1 µm-diameter beads used in some optical trapping studies) has the potential to decrease the probability of multiple-motor transport at a particular motile fraction. For flexible kinesin, however, the impact of bead size on $\alpha$ will depend on how the motor's extension length varies with bead size. We are currently investigating the effect(s) of bead size on kinesin's extension length.

For 500 nm-diameter beads, our data indicate that kinesin extends to 57 nm (~71% of its contour length) in bead-based assays (Fig. 3 *B*). This finding supports the general assumption that kinesin is in an extended conformation during bead transport (for example (3)). However, an extended conformation for kinesin is also surprising, since kinesin was previously found to adopt a more compact conformation in microtubule-gliding assays, extending to only ~28% of its contour length (31). The key difference between bead-based assays and microtubule-gliding assays may be that the interface between microtubules and the cargo surface is highly curved in the former but flat in the latter. Our study raises the intriguing possibility that kinesin can exploit its conformational flexibility to seek and interact with microtubules at highly curved interfaces such as those that occur during vesicle transport in cells.

The experimental approach used here harnesses the increased kinesin/microtubule association time at lower ATP concentrations (14, 15). Thus, this approach is not limited to specific cargo geometry, motor/cargo recruitment methods, or measurement approach (optical trapping or fluorescence). It is, however, specific to studies of kinesin, for which the inverse correlation between multiple-motor travel and ATP level has been directly demonstrated (14-18). Future investigations expanding our ability to tune multiple-motor travel distance for other molecular motors (such as dynein) will critically empower this experimental approach. Despite this limitation, when combined with selective small-molecule inhibitors of other motors (such as dynein (35, 36)), the ATP-based strategy may complement current photobleaching- (37) and force-based (38) methods to shed light on the number of kinesin motors involved in the transport of cargos purified from cellular extracts.

**SUPPORTING MATERIAL**

Supporting text and two supporting figures are attached.

## ACKNOWLEDGMENTS

We thank our reviewers for helpful comments. We thank Steven P. Gross for helpful discussions. We thank Tiffany J. Vora for manuscript editing.

This work was supported by the UC Merced Health Sciences Research Institute Biomedical Seed Grant (to JX), the UC Merced Academic Senate Committee on Research (to JX and AG), the National Institutes of Health (NS048501 to SJK), the National Science Foundation (EF-1038697 to AG), and the James S McDonnell Foundation (to AG).

## AUTHOR CONTRIBUTIONS

JX designed the investigation, QL performed the experiments, AG and JX carried out the theory and simulation work, and SJK purified proteins. All authors analyzed the data. JX, SJK, and AG wrote the manuscript.

# SUPPORTING MATERIAL

**Quantitative Determination of the Probability of Multiple-Motor Transport in Bead-Based Assays**

Qiaochu Li,[1] Stephen J. King,[2] Ajay Gopinathan,[1] and Jing Xu[1, *]

[1]Department of Physics, School of Natural Sciences, University of California, Merced, CA 95343, USA

[2]Burnett School of Biomedical Sciences, University of Central Florida, FL 32827, USA

[*]**Correspondence:** Jing Xu (jxu8@ucmerced.edu)

**RUNNING TITLE:** Probability of Multiple-Motor Transport

# TABLE OF CONTENTS

**SUPPORTING TEXT**

**1. Motile fraction indicates the average number of active motors on a bead**

We used motile fraction as a direct readout for the average number of active motors on a bead (1, 2) (Fig. S2). Motile fraction refers to the probability of beads exhibiting motility along microtubules. While motile-fraction measurements do not require optical trapping, trap-free readout can be difficult to interpret due to the potential presence of dead motors that lack enzymatic activity but can still bind microtubules. When an optical trap is used to confine individual beads to the vicinity of the microtubule (as we did in the current study), the bead must move against the optical trap to demonstrate directed motion along the microtubule. A dead motor may bind the microtubule, but it cannot exert force to drive bead movement against the optical trap, and thus it cannot contribute to a motile event. The resulting motile fraction measurements are therefore not sensitive to the potential effect of dead motors, providing a direct readout for the average number of active motors on the bead.

**2. Travel threshold selection**

When a bead is transported by a single motor, the probability of measuring a travel distance of $x$ μm is described by the single exponential decay $P(x) = e^{-x/d}$, where $d$ is the mean travel distance of a single motor. For the kinesin motor, $d = 1$ μm (1, 3), and the likelihood that single-kinesin travel persists for less than $x_o$ is $1 - e^{-x_0/d}$. We thus expect that 99.9% of beads transported by a single kinesin travel ≤6.9 μm.

**3. Estimations using a previous theory model in reference (2)**

A previous study (2) modeled the probability that a bead is transported by two or more motors as

$$P_{previous}(\geq 2) = \alpha \cdot (1 - e^{-n} - ne^{-n}),$$

where $\alpha$ is the probability that two randomly attached motors on the bead are within simultaneous reach of the microtubule and $n$ is the mean number of active motors available for bead transport. Motile fraction provides a direct experimental readout for $n$ (Supporting Text 1 and Fig. S2).

For kinesin, the previous model estimated that $\alpha = 0.099$, given the bead size employed in this study and in typical optical trapping studies (500 nm diameter) (Fig. S1 *A*). Thus,

$$P_{previous}(\geq 2) = 0.099 \cdot (1 - e^{-n} - ne^{-n}).$$

In our study, we recast the dependence on the mean motor number ($n$) as that on motile fraction, which we measured experimentally, using the relationship *motile fraction* $= 1 - e^{-n}$ (Fig. S2 and (1, 2)). Thus,

$$P_{previous}(\geq 2) = 0.099 \cdot (mf + (1-mf) \cdot \ln(1-mf)),$$

where *mf* denotes the motile fraction. Note that this probability considers all beads, including those that did not interact with the microtubule. We normalized $P_{previous}(\geq 2)$ by the associated motile fraction to determine the probability that a motile event is carried out by two or more motors (magenta line, Fig. 1 *B* and Fig. 4, left).

A similar evaluation for dynein (magenta line, Fig. 4, right) was carried out by substituting $\alpha = 0.039$ into the above expression (Fig. S1 *A*).

## 4. Derivation of Equation 1 in the main text

The probability that a bead is transported by two or more motors is determined by the weighted sum $P(\geq 2) = \sum_{k=2}^{\infty} p(k \mid n) \cdot g(2 \mid k)$, where $n$ is the average number of motors on the bead, $p(k \mid n)$ is the Poisson probability that there are exactly $k$ motors on the bead, and $g(2 \mid k)$ is the probability that at least two of the $k$ motors on the bead are available for transport.

Substituting in $p(k \mid n) = n^k e^{-n} / k!$ and $g(2 \mid k) = 1 - (1-\alpha)^{k-1}$, we obtain

$$P(\geq 2) = \sum_{k=2}^{\infty} \left(\frac{n^k e^{-n}}{k!}\right) \cdot (1 - (1-\alpha)^{k-1})$$

$$= \sum_{k=2}^{\infty} \frac{n^k e^{-n}}{k!} - \sum_{k=2}^{\infty} \frac{n^k e^{-n}(1-\alpha)^{k-1}}{k!}$$

$$= 1 - e^{-n} - ne^{-n} - e^{-n} \sum_{k=2}^{\infty} \frac{n^k (1-\alpha)^{k-1}}{k!}.$$

To derive a closed-form expression for $P(\geq 2)$, we denote the infinite sum in the above expression as $S = \sum_{k=2}^{\infty} \frac{n^k (1-\alpha)^{k-1}}{k!}$. Note that $S = 0$ when $n = 0$.

The derivative of $S$ with respect to $n$ gives rise to

$$\frac{dS}{dn} = \frac{d}{dn}\left(\sum_{k=2}^{\infty} \frac{n^k (1-\alpha)^{k-1}}{k!}\right)$$

$$= \sum_{k=2}^{\infty} \frac{(n(1-\alpha))^{k-1}}{(k-1)!}$$

$$= \sum_{k=1}^{\infty} \frac{(n(1-\alpha))^k}{k!}$$

$$= \sum_{k=0}^{\infty} \frac{(n(1-\alpha))^k}{k!} - 1$$

$$= e^{n(1-\alpha)} - 1.$$

Integrating $\frac{dS}{dn}$ yields $S = \frac{e^{n(1-\alpha)}}{1-\alpha} - n + C$. Since $S = 0$ when $n = 0$, $C = -\frac{1}{1-\alpha}$ and

$$S = \frac{e^{n(1-\alpha)}}{1-\alpha} - n - \frac{1}{1-\alpha}.$$

Substituting the closed form for the infinite sum, we obtain Equation 1 in the main text,

$$P(\geq 2) = 1 + e^{-n}(\frac{\alpha}{1-\alpha}) - \frac{e^{-n\alpha}}{1-\alpha}.$$

## 5. Derivation of Equation 2 in the main text

Our experimental measurements represent lower-bound probabilities of multiple-motor transport, since our distance threshold excludes the population of multiple-motor transport events that travel ≤6.9 μm (Fig. 1 in the main text). Imposing the same threshold, the expected lower-bound probability of multiple-motor transport is

$$P_{lower-bound}(\geq 2) = P(\geq 2) - \sum_{i=2}^{\infty} f_i \cdot P(=i),$$

where $P(\geq 2)$ is the probability of multiple-motor events without imposing any travel threshold (Eq. 1 in the main text), $P(=i)$ is the probability that the bead is transported by exactly $i$ motors, and $f_i$ is the probability that bead travel by exactly $i$ motors exceeds the travel threshold. We previously measured the value of $f_2$ to be 0.556±0.096 for the experimental condition used in the present investigation (0.01 mM ATP) (4). Under the same experimental condition, the value of $f_{i>2}$ approaches 1 based on predictions of multiple-motor travel distances by a theory model in (5). We thus have the following simplification,

$$P_{lower-bound}(\geq 2) = P(\geq 2) - f \cdot P(=2),$$

where $f = f_2 = 0.556$ for simplicity.

The probability that a bead is transported by exactly two motors, $P(=2)$, is determined by

$$P(=2) = \sum_{k=2}^{\infty} p(k|n) \cdot \alpha \cdot (1-\alpha)^{k-2},$$

where $n$ is the average number of motors on the bead, $p(k\,|\,n)$ is the Poisson probability that there are exactly $k$ motors on the bead, $\alpha$ is the probability that two randomly attached motors are within simultaneous reach of the microtubule, and $\alpha \cdot (1-\alpha)^{k-2}$ is the probability that exactly two of the $k$ motors on the bead are available for transport.

We derived a closed-form expression for $P(=2)$ as

$$P(=2) = \sum_{k=2}^{\infty} \frac{n^k e^{-n}}{k!} \cdot \alpha \cdot (1-\alpha)^{k-2}$$

$$= \frac{\alpha \cdot e^{-n}}{1-\alpha} \sum_{k=2}^{\infty} \frac{n^k (1-\alpha)^{k-1}}{k!}.$$

Substituting $\sum_{k=2}^{\infty} \frac{n^k (1-\alpha)^{k-1}}{k!} = S = \frac{e^{n(1-\alpha)}}{1-\alpha} - n - \frac{1}{1-\alpha}$ from Supporting Text 4,

$$P(=2) = \frac{\alpha \cdot e^{-n}}{1-\alpha} \cdot \left(\frac{e^{n-n\alpha}}{1-\alpha} - n - \frac{1}{1-\alpha}\right)$$

$$= \frac{\alpha \cdot e^{-n\alpha}}{(1-\alpha)^2} - \frac{\alpha \cdot n \cdot e^{-n}}{1-\alpha} - \frac{\alpha \cdot e^{-n}}{(1-\alpha)^2}.$$

Thus, $P_{lower-bound}(\geq 2) = P(\geq 2) - f \cdot \left(\frac{\alpha \cdot e^{-n\alpha}}{(1-\alpha)^2} - \frac{\alpha \cdot n \cdot e^{-n}}{1-\alpha} - \frac{\alpha \cdot e^{-n}}{(1-\alpha)^2}\right).$

**SUPPORTING FIGURES**

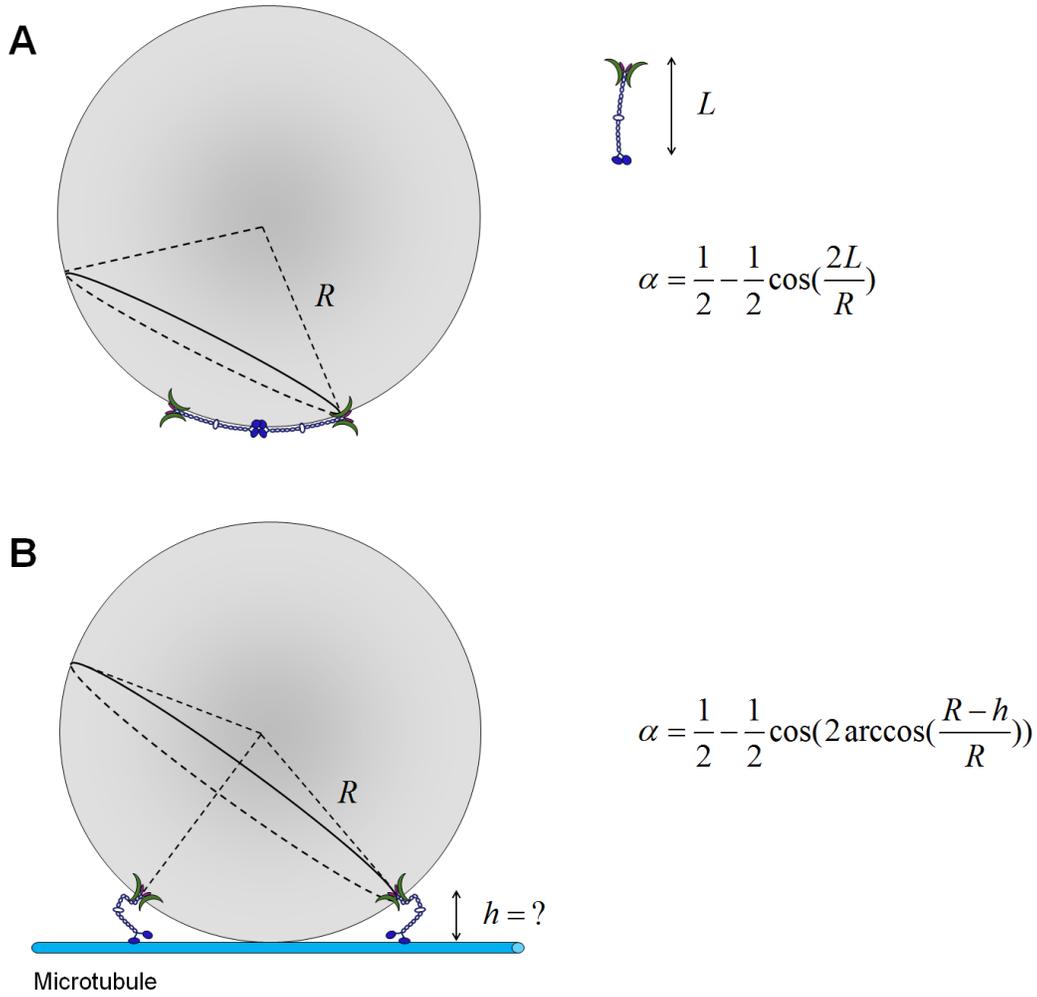

FIGURE S1  Geometries for two-motor transport (not to scale) and associated probability that two randomly attached motors are within simultaneous reach of the microtubule ($\alpha$). Here we considered the condition in which the bead radius ($R$) is larger than the motor's contour length ($L$), and the motors are randomly distributed on the bead surface. Under this condition, although all motors on the bead can contribute to bead transport, not all motors can reach the microtubule at the same time. When any one motor on the bead binds the microtubule, there is a limited area surrounding this motor (a spherical cap) in which a second motor may be located and be within reach of the same microtubule. The value of $\alpha$ is determined as the area ratio of this spherical cap to the entire bead surface. (*A*) Geometry for two-motor transport in a previous model (2). The motors are assumed to be fully extended, and their motor domains effectively in contact with each other. The area of the spherical cap enclosing a second motor available for transport has a radius twice the motor's contour length. $R = 250$ nm in the current study (and in typical optical trapping studies). This geometry yields $\alpha = 0.099$ for kinesin ($L = 80$ nm (6, 7)) and $\alpha = 0.039$

for dynein ($L$ = 50 nm (8, 9)). (*B*) An updated geometry for two-motor transport (10, 11). Here, the motors can bind different locations along the length of the microtubule, and the motors are no longer assumed to be fully extended. The area of the spherical cap enclosing a second motor available for transport is determined by *h*, the extension of the motor during active transport. There is limited information about *h* for flexible motors such as kinesin.

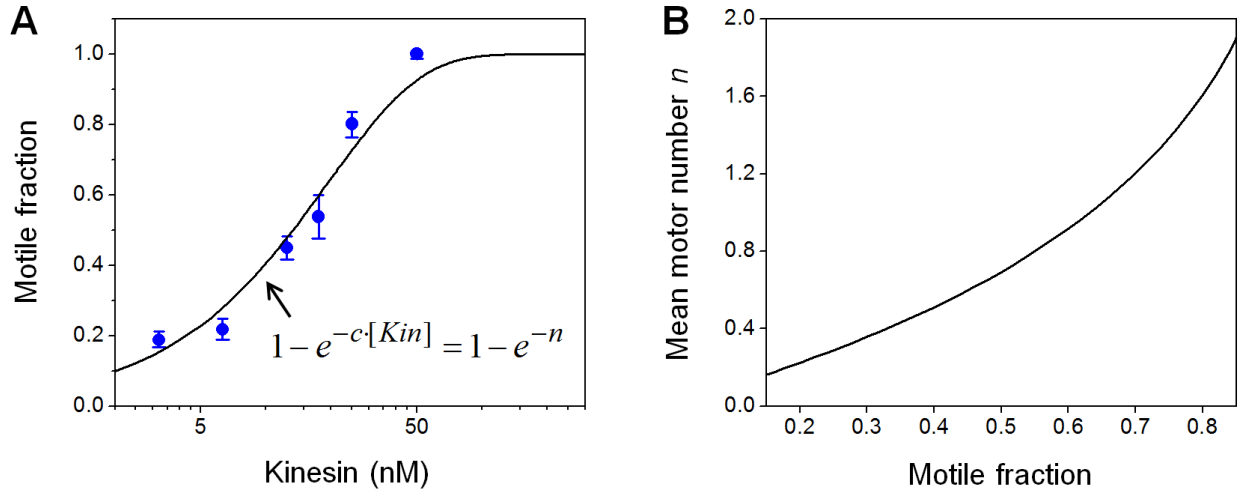

FIGURE S2  Motile fraction provides a direct experimental readout for the average number of active motors on the bead ($n$). (*A*) Fraction of beads exhibiting motility along microtubules ("motile fraction") as a function of kinesin motor concentration, measured using an optical trap and at 1 mM ATP. The concentration of beads was kept constant and the concentration of kinesin was varied. Individual beads were confined to the vicinity of microtubules with an optical trap (80-330 beads trapped for each motor concentration). Error bars represent standard error of the mean. Since motile fraction represents the probability that a bead is carried by at least one active motor, it is well described by the single-motor Poisson curve (1, 2) $P(\geq 1) = 1 - e^{-c \cdot [Kin]} = 1 - e^{-n}$ (solid line), where $c$ represents a fitting parameter and $[Kin]$ indicates kinesin concentration. Note that, we carried out our measurements as a function of motile fraction rather than motor concentration ($[Kin]$), because each motile fraction corresponds to a unique $n$ value. In contrast, for any particular motor concentration ($[Kin]$), the average number of active motors on a bead ($n$) is not unique and depends sensitively on the concentration of beads used during motor/bead incubation (this dependence is reflected in the fitting parameter $c$). (*B*) Relationship between $n$ and motile fraction using *motile fraction* $= 1 - e^{-n}$ (panel *A* and (1, 2)).

**SUPPORTING REFERENCES**